\documentclass[preprint,showpacs,preprintnumbers,amsmath,amssymb]{revtex4}

\usepackage{graphicx}
\usepackage{dcolumn}
\usepackage{bm}
\usepackage{psfig}

\begin{document}

\preprint{APS/123-QED}

\title{Interferometric detection of spin-polarized transport in the depletion layer of a metal-GaAs Schottky barrier}

\author{G. Salis}
\email{gsa@zurich.ibm.com}
\author {S. F. Alvarado}
\affiliation{IBM Research, Zurich Research Laboratory,
S\"aumerstrasse 4, 8803 R\"uschlikon, Switzerland}

\date{February 8, 2006}

\begin{abstract}
It is shown that the Kerr rotation of spin-polarized electrons is
modulated by the distance of the electrons from the sample
surface. Time-resolved Kerr rotation of optically-excited
spin-polarized electrons in the depletion layer of $n$-doped GaAs
displays fast oscillations that originate from an interference
between the light reflected from the semiconductor surface and
from the front of the electron distribution moving into the
semiconductor. Using this effect, the dynamics of the
photogenerated charge carriers in the depletion layer of the
biased Schottky barrier is measured.
\end{abstract}


\maketitle

Time-resolved optical techniques allow the charge and spin
dynamics in semiconductors to be tracked on ultrashort time scales
and with a spatial resolution that is only limited by the optical
wavelength~\cite{Shah99}. The charge of photoexcited electrons and
holes generates an electric field that is observable by
differential transmission or reflection measurements
~\cite{Shank1981,Dekorsy1993,Achermann2002}. Information on spin
dynamics can be obtained by the magneto-optical Kerr and Faraday
rotation~\cite{Kikkawa1999,Kato2004,Crooker2005}. In the presence
of a magnetization in the sample, linearly polarized probe light
is rotated by an angle $\theta$ after reflection off the sample
(Kerr) or transmission through the sample (Faraday). In the polar
geometry, $\theta$ is assumed to be proportional to the spin
polarization along the probe beam.

Here we show that in addition to spin polarization, Kerr rotation
(KR) is highly sensitive to the charge-carrier distribution of the
spin-polarized electrons. The reason for this is an optical
interference of the probe beam reflected off the spin-polarized
electron distribution and of that reflected directly off the
sample surface. This leads to an oscillation of the KR amplitude
with the distance of the spin-polarized electrons from the sample
surface, and makes it possible to investigate perpendicular
transport of spin-polarized electrons with high sensitivity and
spatial resolution. The optical interference effect discussed here
is similar to the observed KR dependence on the thicknesses of
metallic magnetic layers~\cite{Katayama1988}, but unrelated to
quantum size effects found in metallic layers~\cite{Suzuki1998}.
We demonstrate the effect for spin-polarized electrons optically
excited in the depletion layer of a Schottky barrier in $n$-doped
GaAs. The electric field in the depletion layer pushes the
electrons into the semiconductor, leading to an oscillation of the
time-resolved KR signal that can be clearly separated from spin
precession induced by an applied magnetic field. From the
oscillations we deduce the position of the receding electron front
as a function of time, and find a strong screening by the
space-charge field of the injected electrons and holes. When the
electrons reach the end of the depletion layer, the KR oscillation
stops, and the thickness of the depletion layer can be determined
from the number of oscillations. We find good agreement with the
expected layer thickness, which we vary by applying a bias across
the Schottky barrier and by changing the doping concentration of
the sample.

\begin{figure}
\includegraphics[width=80mm]{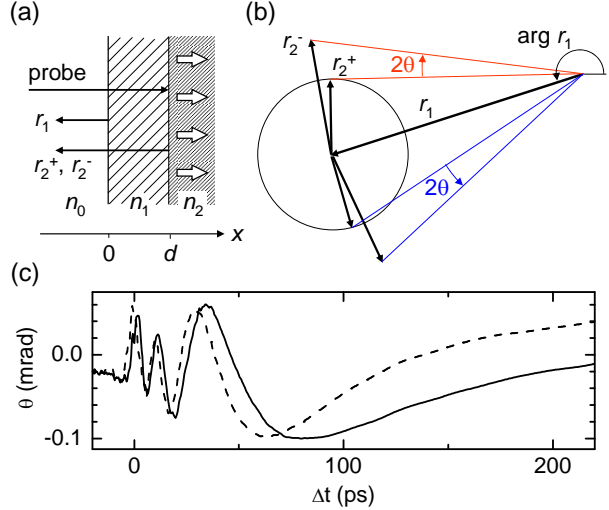}
\caption{\label{fig:fig1} (Color online) (a) Scheme of a sample
with a spacer layer ($n_1$) and a spin-polarized layer ($n_2$). A
probe beam is reflected at the sample surface ($r_1$) and at the
boundary to the spin-polarized layer ($r_2^+$ and $r_2^-$). (b)
Complex plane representation of reflection amplitudes and KR angle
$\theta$. (c) Measured $\theta(\Delta t)$ in $n$-doped GaAs
(sample\,1) with an applied bias of --1.0\,V (solid line) and
--1.1\,V (dashed line) across the Schottky barrier. Oscillations
in $\theta$ are due to the displacement of the spin-polarized
electron front into the semiconductor.}
\end{figure}

We first discuss how the KR angle $\theta$ is related to the
spin-induced circular birefringence (CB) and circular dichroism
(CD) of a spin-polarized region that starts at a distance $d$ from
the sample surface, and show that the amplitude of $\theta$
oscillates with $d$. We assume that in a sample with its surface
at $x=0$, spin-polarized electrons are present only for $x>d$,
with a spacer layer at $0<x<d$, see Fig.~\ref{fig:fig1}(a). The
refractive index of the spacer layer is $n_1$, and that of the
spin-polarized layer $n_2^+$ or $n_2^-$, depending on the helicity
of the photon. CB (CD) leads to a helicity-dependence of the real
(imaginary) part of $n_2$. In semiconductors, CB and CD originate
from a spin-dependent state-filling effect combined with optical
selection rules~\cite{Snelling1994}. The Kerr rotation angle
$\theta$ is given by $(\arg r^+ - \arg r^-)/2$, where $\arg
r^{\pm}$ is the phase of the helicity-dependent total reflection
amplitude $r^{\pm}=r_1+r_2^{\pm}$ at wavelength $\lambda$. The
reflection amplitude at the surface is $r_1=(n_0-n_1)/(n_0+n_1)$,
where $n_0$ is the refractive index at $x<0$. The amplitude $r_2$
describes the reflection at the interface to the spin-polarized
layer, including transmission through the surface and propagation
in the spacer layer. We write $r_2^{\pm}=|t_{01}t_{10}|e^{i(
\phi_0+\phi)}(n_1-n_2^{\pm})/(n_1+n_2^{\pm})$, with $\phi=4\pi n_1
d/\lambda$, $\phi_0=\arg t_{01}t_{10}$, and $t_{01}$ ($t_{10}$)
being the transmission amplitudes through the surface from left to
right (right to left).

Figure~\ref{fig:fig1}(b) schematically shows the reflection
amplitudes in the complex plane. Because of CB and CD, $r_2^+$
differs in phase and amplitude from $r_2^-$, but the differences
do not depend on $d$ and therefore on $\phi$. As $\phi$ is varied,
$r_2^+$ and $r_2^-$ rotate in the complex plane, leading to an
oscillation of $\theta$ with $\phi$ (harmonic in $\phi$ for
$|r_1|>>|r_2^{\pm}|$). Peak values in $\theta$ of opposite sign
occur at values that differ in $\phi$ by $\pi$, as shown in
Fig.~\ref{fig:fig1}(b). In the following, we experimentally
demonstrate the occurrence of such KR modulation by varying $d$,
and show that this effect can be used for time-resolved
measurement of the electron dynamics in a Schottky-barrier
depletion layer.

Samples were prepared from $n$-doped GaAs wafers with nominal
doping concentrations $\rho$ of 1$\times$10$^{16}$ (sample 1) and
8$\times$10$^{16}$\,cm$^{-3}$ (sample 2), providing long
spin-coherence times~\cite{Kikkawa1998}. A 1.5\,nm thick
Al$_2$O$_3$ barrier, a 5.8\,nm thick Al layer and a 4\,nm thick Au
capping layer were thermally evaporated onto the (001) surface of
the wafers, serving as a semitransparent electrode. A bias $U$ was
applied between the electrode and an ohmic contact on the back
side of the samples (with a negative $U$ corresponding to the
negative pole on the electrode side). Circularly polarized,
2--3\,ps long pump pulses from a mode-locked Ti:sapphire laser are
used to excite spin-polarized electron-hole pairs in the
semiconductor. The KR angle $\theta(\Delta t)$ is detected of
linearly polarized probe pulses that are delayed with respect to
the pump pulses by a time $\Delta t$, see~\cite{Salis2005} for
further details. Both beams are tuned to the GaAs absorption edge
at $\lambda=819.8$\,nm and focused onto a $\approx$30$\,\mu$m wide
spot on the electrode. If not noted otherwise, the pump power is
650\,$\mu$W and the probe power 70\,$\mu$W. The samples were
placed in an optical cryostat and cooled to 10\,K. Using a
transfer-matrix calculation, we obtain a transmission coefficient
through the electrode of 22\%, which yields an average
concentration of optically excited charge carriers of
10$^{12}$\,cm$^{-2}$.

\begin{figure}
\includegraphics[width=85mm]{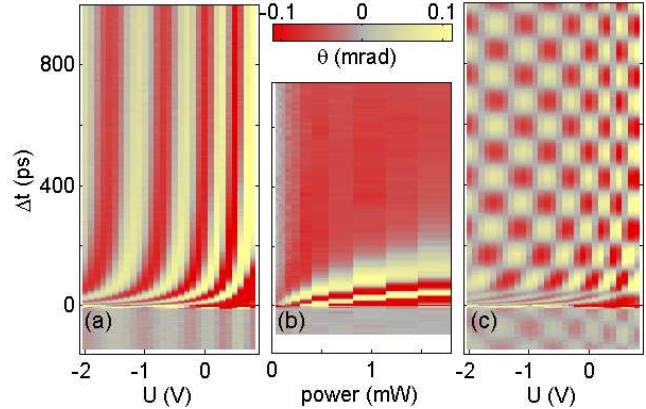}
\caption{\label{fig:fig2} (Color online) Measured $\theta$ on
sample 1. (a) Map of $\theta$ vs. $\Delta t$ and $U$ at
$B$\,=\,0\,T and 650\,$\mu$W pump power. (b) Dependence of
$\theta$ vs. $\Delta t$ on pump power at $U=0$\,V. (c): same as
(a) with $B$\,=\,1\,T.}
\end{figure}

Figure~\ref{fig:fig1}(c) shows KR traces $\theta(\Delta t)$ for
sample\,1 at $U=-1.0$ and $-1.1$\,V, exhibiting initially fast
oscillations at $\Delta t>0$ whose frequency decreases at higher
$\Delta t$. For more negative $U$, $\theta$ oscillates faster. In
Fig.~\ref{fig:fig2}(a), the dependency of $\theta$ on $\Delta t$
and $U$ is presented as a density plot. We see that for a given
$U$, $\theta$ reaches an asymptotic value at $\Delta t>0.5$\,ns
that is determined by and oscillates with $U$. Between --2.0 and
0.8\,V, about four full oscillations can be resolved.

The oscillations in $\theta$ with $U$ and $\Delta t$ can be
explained in the context of the optical interference discussed
above: The pump pulses create spin-polarized charge carriers that
extend from the semiconductor surface about 2--10\,$\mu$m into the
semiconductor. Driven by the strong electric field in the
depletion layer, the spin-polarized electrons will drift into the
semiconductor, whereas the holes are pushed towards the electrode.
As the distance $d$ between the front of the spin-polarized
electrons and the semiconductor surface increases, $\phi$
increases and therefore $\theta$ oscillates. As the electron front
moves into the semiconductor, the electric field weakens and the
drift slows down until the electrons reach the end of the
depletion layer at $d=d_{\rm{depl}}$, where the phase of the
oscillations attains an asymptotic value that linearly depends on
$d_{\rm{depl}}$. Since $d_{\rm{depl}}$ can be varied by changing
$U$, the asymptotic value of $\theta$ oscillates with $U$.

The electric field in the depletion layer is dynamically screened
by the photoexcited electron-hole pairs. A variation of the pump
power should affect the screening and thus the dynamics $d(\Delta
t)$ of the moving electron front, but not its end position at
$d_{\rm{depl}}$. This is confirmed in a measurement of the
pump-power dependence of $\theta(\Delta t)$ at $U=0$\,V
[Fig.~\ref{fig:fig2}(b)], clearly showing that although the time
spacing of the oscillations varies widely with pump power, the
asymptotic value at long $\Delta t$ is not affected. Dynamic
screening directly affects the dielectric function through the
Franz-Keldysh effect~\cite{Shank1981}, which could in principle
also lead to oscillations in $\theta(U,\Delta t)$. However such
oscillations would neither exhibit the observed dependence on
$d_{\rm{depl}}$\cite{remark3} nor would its asymptotic phase be
independent of pump power\cite{remark4}.

Before discussing the interference mechanism further, we exclude
in the following that the oscillations in $\theta$ are induced by
the dynamics of the electron spin. For instance, the oscillations
in $\theta$ could be induced by spin precession about an internal
magnetic field due to hyperfine interaction with nuclear spins or
spin-orbit effects. If this were the case, an external magnetic
field $B$ would either enhance or decrease the period of the
oscillations of $\theta$. In Fig.~\ref{fig:fig2}(c),
$\theta(\Delta t,U)$ is shown with $B=1.0$\,T applied in the (110)
direction of the GaAs wafer. Superposed onto the same modulation
of $\theta$ as at $B=0$, $\theta$ now exhibits a periodic
oscillation in $\Delta t$ due to spin precession about $B$. Also,
a magnetic field applied along the (100) direction does not affect
the $B=0$ modulation of $\theta(\Delta t,U)$ (data not shown). In
fact, $\theta$ can be written as the product of a component
proportional to the electron spin polarization and a
spin-independent factor that oscillates with $\Delta t$ and $U$.

\begin{figure}
\includegraphics{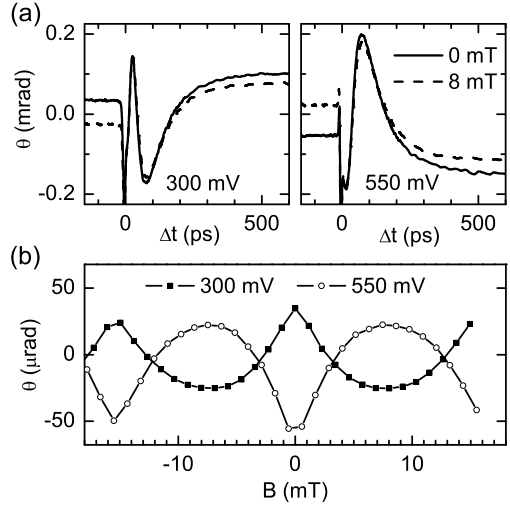}
\caption{\label{fig:fig3} (a) Measured $\theta$ vs. $\Delta t$ at
$B=0$ (solid lines) and 8\,mT (dashed lines), and for $U=300$ and
550\,mV (sample 1). In (b), $\theta$ vs. $B$ at $\Delta
t$\,=\,12.4\,ns shows resonant spin amplification at both $U=300$
and 550\,mV, with opposite signs for the two cases.}
\end{figure}

Further insight into the origin of the $\theta$ modulation is
obtained by studying how spins excited by an earlier laser pulse
sum up with those of a new pulse. Every 12.5\,ns, a pump pulse
excites a new spin packet. If the spins of an old packet have
precessed an integer number of cycles until a new packet is
excited, the KR signal is resonantly amplified~\cite{Kikkawa1998}.
If the observed modulation of $\theta$ were due to spin
precession, different amplifications would be expected for
negative and positive $\theta$ upon arrival of a new pump pulse. A
positive correlation would give rise to a resonant build-up, a
negative one would suppress the resonance. As seen in
Fig.~\ref{fig:fig3}(a), $\theta(\Delta t)$ saturates at a positive
(negative) value for $U=300$\,mV (550\,mV), $B=0$\,T and $\Delta
t=12.5$\,ns (equivalent to $\Delta t=0$). For $B=8$\,mT, the spins
have precessed half a cycle, and $\theta$ has reversed its sign at
$\Delta t=12.5$\,ns. In Fig.~\ref{fig:fig3}(b), the $B$-dependence
of $\theta$ is shown at $\Delta t=12.4$\,ns and for 300 and
550\,mV. The peak at 0\,mT is repeated at --15.5\,mT,
corresponding to one full precession cycle in 12.5\,ns. Reversing
the sign of $\theta$ at 12.5\,ns by varying $U$ does not suppress
resonant spin amplification, but only reverses its sign. This is a
clear indication that the modulations of $\theta$ at $B=0$ are not
related to spin precession.

Figure~\ref{fig:fig3}(a) reveals the dependence of the KR signal
on the spatial distribution of the electron spins: Immediately
after injection of the new spin packet at $\Delta t=0$, $\theta$
does not depend on whether the old spin packet has precessed half
a turn (8\,mT) or not (0\,mT)~\cite{remark1}. As the polarization
of the old spin packet has not decayed completely, the newly
excited spins must mask the old spins. This is possible because
the new spin packet has a different spatial position that starts
directly at the GaAs surface, whereas the old spins have already
travelled behind the depletion layer. As long as the new spins are
closer to the semiconductor surface than the old spins, they
predominate the KR signal. The old spins will contribute to the
signal and possibly resonantly amplify it once the new spins have
arrived behind the depletion layer and mix with the old ones.

Assuming that the modulation in $\theta$ is due to an interference
of the reflected beams from the metallic electrode and the front
of the spin-polarized electrons at position $d$, we obtain
$d(\Delta t)$ from the data shown in Fig.~\ref{fig:fig2}(a). After
numbering subsequent peaks (both positive and negative ones) in
$\theta(\Delta t)$ by an integer $k$, we obtain
$d=d_0+k\lambda/4n_1$, where $d_0$ is an offset that depends on
several quantities: The phase change of reflection and
transmission at the metallic electrode, the helicity-dependent
reflection at the receding front of the spin-polarized electrons,
a smearing of the front due to carrier diffusion, and a correct
numbering of the interferences. Using a transfer-matrix approach
with literature values for the refractive indices of Au, Al,
Al$_2$O$_3$ and GaAs, we calculate $\phi_0\approx-0.24\pi$ and
$\arg r_1\approx -0.92\pi$. Surprisingly, the phases $\phi$ for
which peaks in $\theta$ occur depend only on the relative strength
of CD and CB, i.e. on the ratio of the real and complex parts of
$n_2^+ - n_2^-$, and not on $n_1-n_2^{\pm}$. This is because the
contrast between $n_1$ and $n_2^{\pm}$ not only influences the
phase of $r_2^{\pm}$, but also the difference in $r_2^+$ and
$r_2^-$. Both contributions compensate each other in their effect
on the position $\phi$ of maximum $\theta$. Peaks in $\theta$ are
expected at $\phi=\arg r_1-\phi_0+k\pi$ for CD and at $\phi=\arg
r_1-\phi_0+\pi/2+k\pi$ for CB. From this and assuming $n_1=3.56$,
we obtain $d_0$\,=\,--39 (--10)\,nm for predominantly CD (CB).
After $\theta$ becomes negative at $U\approx500$\,mV and $\Delta
t=50$\,ps, no new peak emerges for higher $U$
[Fig.~\ref{fig:fig2}(a)]. This suggests that this negative value
corresponds to $k=1$. At more negative $U$, the initial
oscillations start to smear out because of finite temporal
resolution, and the correct numbering is obtained by extrapolation
to data at positive $U$.

\begin{figure}
\includegraphics{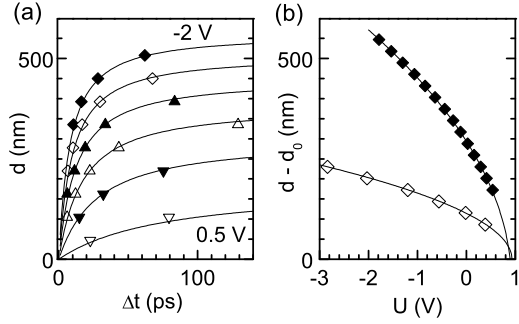}
\caption{\label{fig:fig4} (a) Measured $d$ vs. $\Delta t$ for
sample 1 and 650\,$\mu$W pump power, with $U$ varying between -2
and 0.5\,V in 0.5\,V steps; lines are fits as described in the
text. In (b), $d-d_0$ vs. $U$ at $\Delta t=2.3$\,ns is plotted for
sample 1 (solid diamonds) and 2 (open diamonds) together with fits
of the depletion-layer thickness $d_{\rm depl}$ (lines).}
\end{figure}

In Fig.~\ref{fig:fig4}(a), $d(\Delta t)$ is plotted for sample 1
(symbols) measured at a pump power of 650\,$\mu$W and taking
$d_0$\,=\,--10\,nm. The data is fitted using
$d(t)=d_{\rm{depl}}(1-(1+zv_0t/d_{\rm{depl}})^{-1/z})$, which is
derived from $\delta d/\delta t=v_0 (1-x/d_{\rm{depl}})^{1+z}$,
where $v_0$ is the velocity of the electron front at $\Delta t=0$.
The latter is motivated by assuming a screening of the electric
field in the depletion layer, $E_0(1-x/d_{\rm{depl}})$,
proportional to $(1-x/d_{\rm{depl}})^z$. Without screening, $z=0$,
and $d=d_{\rm{depl}}(1-\exp(-v_0 t/d_{\rm{depl}}))$. The data in
Fig.~\ref{fig:fig4}(a) are fitted by expressing $E_0$ and
$d_{\rm{depl}}$ by $U$:
$E_0=\sqrt{2e\rho(U_0-U)/\epsilon\epsilon_0}$ and
$d_{\rm{depl}}=\epsilon\epsilon_0E_0/\rho e$. $U_0$ is the
built-in potential across the Schottky barrier and $\epsilon=13$
the dielectric constant of GaAs. We find a dependence of $v_0$ on
$E_0$ that can be approximated as $v_0=\alpha E_0^3$, with
$\alpha=6\times10^{-9}$\,cm$^4$/V$^3$s. We obtain $z$\,=\,0.75,
$U_0$\,=0.79\,V, and $\rho=1.3\times10^{16}$\,cm$^{-3}$. For
$U=0$\,V, we therefore have $E_0=5.4\times10^{4}$\,V/cm,
$d_{\rm{depl}}=300$\,nm, and $v_0=9.4\times10^5$\,cm/s. A fit with
$d_0=-39$\,nm yields similar values, but has a higher standard
deviation. The velocity $v_0$ is about 10 times smaller than the
high-field drift velocity in GaAs~\cite{Blakemore1982} and about
$400$ times smaller than $\mu E_0$, with $\mu=8000$\,cm$^2$/Vs.
These results are evidence of a space-charge-related origin of the
observed dynamics, in which the spatially separating electrons and
holes screen the electric field~\cite{Dekorsy1993,Achermann2002}.
The screening is limited by the excited charge density, which
explains the observed superlinear increase of $v_0$ with $E_0$. As
can be seen in Fig.~\ref{fig:fig2}(b), the screening strongly
decreases as the pump power and thus the density of excited
electron-hole pairs is reduced.

By numbering the oscillations of $\theta(U)$ at constant $\Delta
t$, we obtain $d-d_0$ vs. $U$. At $\Delta t=2.3$\,ns, $d\approx
d_{\rm{depl}}$, and $d_{\rm{depl}}(U)$ can be obtained
[Fig.~\ref{fig:fig4}(b)]. The data is in excellent agreement with
a fit to the analytical expression for $d_{\rm{depl}}(U)$, for
which we obtain $\rho=1.1\times10^{16}$
($1.0\times10^{17}$)\,cm$^{-3}$, $U_0=0.92$ (0.93)\,V, and
$d_0$=-60 (0)\,nm for sample 1 (sample 2). The fitted $\rho$ is
slightly higher than the nominal doping concentration, which might
be related to the electron front not being completely pushed out
of the depletion layer, or to a smaller dielectric constant than
the assumed values ($\epsilon=13$, $n=3.56$~\cite{Blakemore1982}).

We find evidence of the interference effect also in GaAs without
metallization. Note that for the position-modulated KR to occur,
the only spin-sensitive contribution to the reflected probe beam
must be the reflection at the electron front. Specifically, no
modulation is expected in a Faraday rotation geometry and for thin
epilayers where an additional reflection from the sample's back
side interferes with the measured beam.

In conclusion, we have observed that the KR signal of
spin-polarized electrons in bulk $n$-doped GaAs oscillates after
optical excitation of spin-polarized electron-hole pairs. This is
explained by an optical interference of the weak reflection at the
receding front of the spin-polarized electron distribution with
the strong reflection at the semiconductor surface. This new
method allows one to directly map the position of spin-polarized
electrons on a picosecond time scale and to characterize
semiconductor properties of the bulk and of
interfaces~\cite{remark2}.

We acknowledge M. Tschudy and M. Witzig for technical support, as
well as R. Allenspach, R. Mahrt and L. Meier for fruitful
discussions.

\newpage


\begin{thebibliography}{17}
\bibitem[{rem({\natexlab{b}})}]{remark4}
\bibinfo{note}{When the electrons arrive at the end of the depletion layer, a
  pump-power dependent screening would still affect the phase of Franz-Keldysh
  oscillations until all excess electrons have recombined.}
\bibitem[{rem({\natexlab{c}})}]{remark1}
\bibinfo{note}{The decreasing amplitude of $\theta$ with $\Delta t$ at 8\,mT
  can be explained by the slow precession of the spin.}
\bibitem[{rem({\natexlab{d}})}]{remark2}
\bibinfo{note}{Also the total reflectivity is modulated by $\phi$, which is
  measurable as an oscillation in differential reflectivity and makes it
  possible to map the position of unpolarized electrons.}
\end{thebibliography}

\end{document}